\documentclass[times, twocolumn, trackchanges]{aastex631} 
\usepackage{graphicx, graphics}
\usepackage{CJK}
\usepackage{amsmath, amssymb, bm}
\usepackage[nodayofweek]{datetime}
\usepackage{multirow}
\input{prepcitationformat.txt}

\newenvironment{rotatepage}%
    {\clearpage\pagebreak[4]\global\pdfpageattr\expandafter{\the\pdfpageattr/Rotate 90}}%
    {\clearpage\pagebreak[4]\global\pdfpageattr\expandafter{\the\pdfpageattr/Rotate 0}}
\makeatletter
\newcommand*{\centerfloat}{%
  \parindent \z@
  \leftskip \z@ \@plus 1fil \@minus \textwidth
  \rightskip\leftskip
  \parfillskip \z@skip}
\makeatother

\newdateformat{monthyearday}{%
  \THEYEAR\ \monthname[\THEMONTH] \twodigit{\THEDAY}}

\newcommand{\Qphi}{$\mathcal{Q}_\phi$}

\newcommand{\uat}[2]{\href{http://vocabs.ands.org.au/repository/api/lda/aas/the-unified-astronomy-thesaurus/current/resource.html?uri=http://astrothesaurus.org/uat/#1}{#2 (#1)}}

\newcommand{\affilCaltechAstro}{\affiliation{Department of Astronomy, California Institute of Technology, MC 249-17, 1200 East California Boulevard, Pasadena, CA 91125, USA; \url{ren@caltech.edu}}}

\received{2021 October 4}
\revised{2021 November 23} 
\accepted{2021 November 24}
\acceptjournal{The Astrophysical Journal Letters}

\shorttitle{HD~34282 Keck/NIRC2 Imaging}
\shortauthors{Quiroz et al.}

\begin{document}
\pagenumbering{arabic}
\begin{CJK*}{UTF8}{gbsn}
\title{Improving Planet Detection with Disk Modeling:\\ Keck/NIRC2 Imaging of the HD~34282 Single-armed Protoplanetary Disk}

\author[0000-0001-8037-3779]{Juan Quiroz}
\affilCaltechAstro

\author[0000-0003-0354-0187]{Nicole L. Wallack}
\affiliation{Division of Geological \& Planetary Sciences, California Institute of Technology, MC 170-25, 1200 East California Boulevard, Pasadena, CA 91125, USA; \url{nwallack@caltech.edu}}

\author[0000-0003-1698-9696]{Bin Ren (任彬)}
\altaffiliation{To whom correspondence should be addressed.}
\affilCaltechAstro

\author[0000-0001-9290-7846]{Ruobing Dong  (董若冰)}
\affiliation{Department of Physics \& Astronomy, University of Victoria, Victoria, BC, V8P 1A1, Canada}

\author[0000-0002-6618-1137]{Jerry W. Xuan}
\affilCaltechAstro

\author[0000-0002-8895-4735]{Dimitri Mawet}
\affilCaltechAstro
\affiliation{Jet Propulsion Laboratory, California Institute of Technology, 4800 Oak Grove Drive, Pasadena, CA 91109, USA}

\author[0000-0001-6205-9233]{Maxwell A. Millar-Blanchaer}
\affiliation{Department of Physics, University of California, Santa Barbara, CA 93106, USA}

\author[0000-0003-4769-1665]{Garreth Ruane}
\affiliation{Jet Propulsion Laboratory, California Institute of Technology, 4800 Oak Grove Drive, Pasadena, CA 91109, USA}

\begin{abstract}
Formed in protoplanetary disks around young stars, giant planets can leave observational features such as spirals and gaps in their natal disks through planet-disk interactions. Although such features can indicate the existence of giant planets, protoplanetary disk signals can overwhelm the innate luminosity of planets. Therefore, in order to image planets that are embedded in disks, it is necessary to remove the contamination from the disks to reveal the planets possibly hiding within their natal environments. We observe and directly model the detected disk in the Keck/NIRC2 vortex coronagraph $L'$-band observations of the single-armed protoplanetary disk around HD~34282.  Despite a non-detection of companions for HD~34282, this direct disk modeling improves planet detection sensitivity by up to a factor of 2 in flux ratio and ${\sim}10~M_{\rm Jupiter}$ in mass. This suggests that performing disk modeling can improve directly imaged planet detection limits in systems with visible scattered light disks, and can help to better constrain the occurrence rates of self-luminous planets in these systems.
\end{abstract}

\keywords{\uat{1300}{Protoplanetary disks}; \uat{313}{Coronagraphic imaging}; \uat{1257}{Planetary system formation}}

\section{Introduction}
The most recent generation high-contrast imaging surveys that utilize extreme adaptive optics systems have obtained an occurrence rate of ${\lesssim}10\%$ for young self-luminous giant planets \citep[i.e.,][]{nielsen19, vigan21}. Despite their low direct occurrence rates using contemporary instruments,  giant planets can gravitationally interact with their surrounding gaseous protoplanetary disks and leave their mark as observational signatures in the disk structure \citep[e.g., spirals, gaps:][]{dong15spiralPlanet, dong15gapPlanet, bae18}. Therefore, protoplanetary disks with suspected embedded forming planets not only are excellent targets for giant planet detection \citep[e.g.,][]{haffert19, wang20}, but also can help constrain planet-disk interactions \citep[e.g.,][]{bae19, rosotti20}.

While protoplanetary disks with planet-disk interaction signatures may be prime targets for observing actively forming planets, they can also hinder the detection of planets \citep[cf.][]{keppler18, wang20}. Specifically, using the current prevailing observation and data reduction strategy for ground-based high-contrast imaging (i.e., angular differential imaging, or ADI; \citealp{marois06}), disk signals can remain in the reduced images and overwhelm planetary signals, resulting in lower sensitivity to detecting planets in the regions where disk signals dominate (e.g., Figure~12 of \citealp{maire17}, Figure~6 of \citealp{deBoer21}, Figure~4 of \citealp{AsensioTorres21}). Moreover, even after processing, disk signals can mimic the appearances of protoplanets (e.g., HD~100546: \citealp{quanz13, currie14}; HD~169142: \citealp{reggiani14, biller14}), making it challenging to distinguish planets from disk features. Indeed, it is only after modeling and removing the disk from the PDS~70 system that \citet{wang20} could recover the disk-embedded protoplanet planet PDS~70c. Therefore, in addition to utilizing diverse approaches in both observation and data reduction methods, it is necessary to investigate whether disk modeling could also improve our sensitivity to planets detected via ground-based high-contrast imaging, which could possibly lead to more detections in blind-search surveys.

Although spirals and rings are indications of the existence of planets, they can also possibly result from mechanisms that do not involve planets (e.g., spirals: \citealp{dong15spiralGI, montesinos18, hall20}, gaps: \citealp{birnstiel15, vandermarel18}). Although the formation of spirals in protoplanetary disks can be caused both with and without planets, the differing mechanisms may be distrinugishable via the structure in the disk. For example, gravitational instability triggered spiral arms are likely symmetric and have even-numbered spirals \citep[e.g.,][]{dong15spiralGI}, whereas planet-induced spirals can be less symmetric. Another reason to search spiral disks for planet signals is that gap-opening planets \citep[e.g.,][]{zhang18, lodato19} could be less massive than spiral-arm-driving planets \citep[e.g., Figure~1 of][]{bae18}, and most gap opening planets are beyond the detection limits of current instruments. Therefore, in order to have the best chance of detecting a massive companion, herein we focus on planet detection in the single spiral arm system HD~34282 \citep{deBoer21}. Specifically focusing on improving planetary detection limits by modeling of the observed disk structure.

Located at $309\pm2$~pc \citep{gaiaedr3}, HD~34282 is an A3V star \citep{merin04} with a protoplanetary disk containing a 75~au cavity\footnote{Size scaled to match \citet{gaiaedr3} distance.} in $0.87$~mm ALMA continuum emission observations \citep[][]{vanderPlas17}. With an overdensity in the south-east region of its ALMA detected ring which has an inclination of $59\fdg3\pm0\fdg4$ and a position angle of $117\fdg1\pm0\fdg3$, \citet{vanderPlas17} suggested that a ${\approx}50~M_{\rm Jupiter}$ brown dwarf companion at ${\approx}0\farcs1$ could be responsible for shepherding the dust in the HD~34282 system, favoring this explanation over a non-planet driven photoevaporation mechanism for opening the cavity. In $J$-band polarized light observations using VLT/SPHERE, \citet{deBoer21} identified two rings with a ${\sim}56^\circ$ inclination and a ${\sim}119^\circ$ position angle, and a possible tightly-wound single-arm spiral which resembles the pattern driven by a sub-Jupiter to Jupiter mass planet \citep[e.g.,][]{dong15gapPlanet, dong17}. However, the observations in \citet{deBoer21} were not sensitive enough to detect a planet of such mass. Moreover, the planet detection map in Figure~6 of \citet{deBoer21} is affected by the signal of the disk, which causes the the sensitivity to drop from  ${\lesssim}4~M_{\rm Jupiter}$ in exterior regions of the image to ${\gtrsim}10~M_{\rm Jupiter}$ in the regions hosting disk signals.

Herein we present $L'$-band imaging of the HD~34282 system using the Keck/NIRC2 vortex coronagraph. Motivated by the recovery of the PDS~70c planet via disk modeling in \citet{wang20}, we investigate the affect of disk modeling in planet detection using HD~34282. Specifically, we model the observed NIRC2 disk as an experiment in a targeted search of planets that are embedded in disks. We describe the NIRC2 observations in Section~\ref{sec-obs}, we present the observed features and investigate the effects of disk modeling in Section~\ref{sec-ana}, and we summarize our findings in Section~\ref{sec-sum}.

\section{Observation and Data Reduction}\label{sec-obs}

\begin{figure*}[htb!]
	\includegraphics[width=\textwidth]{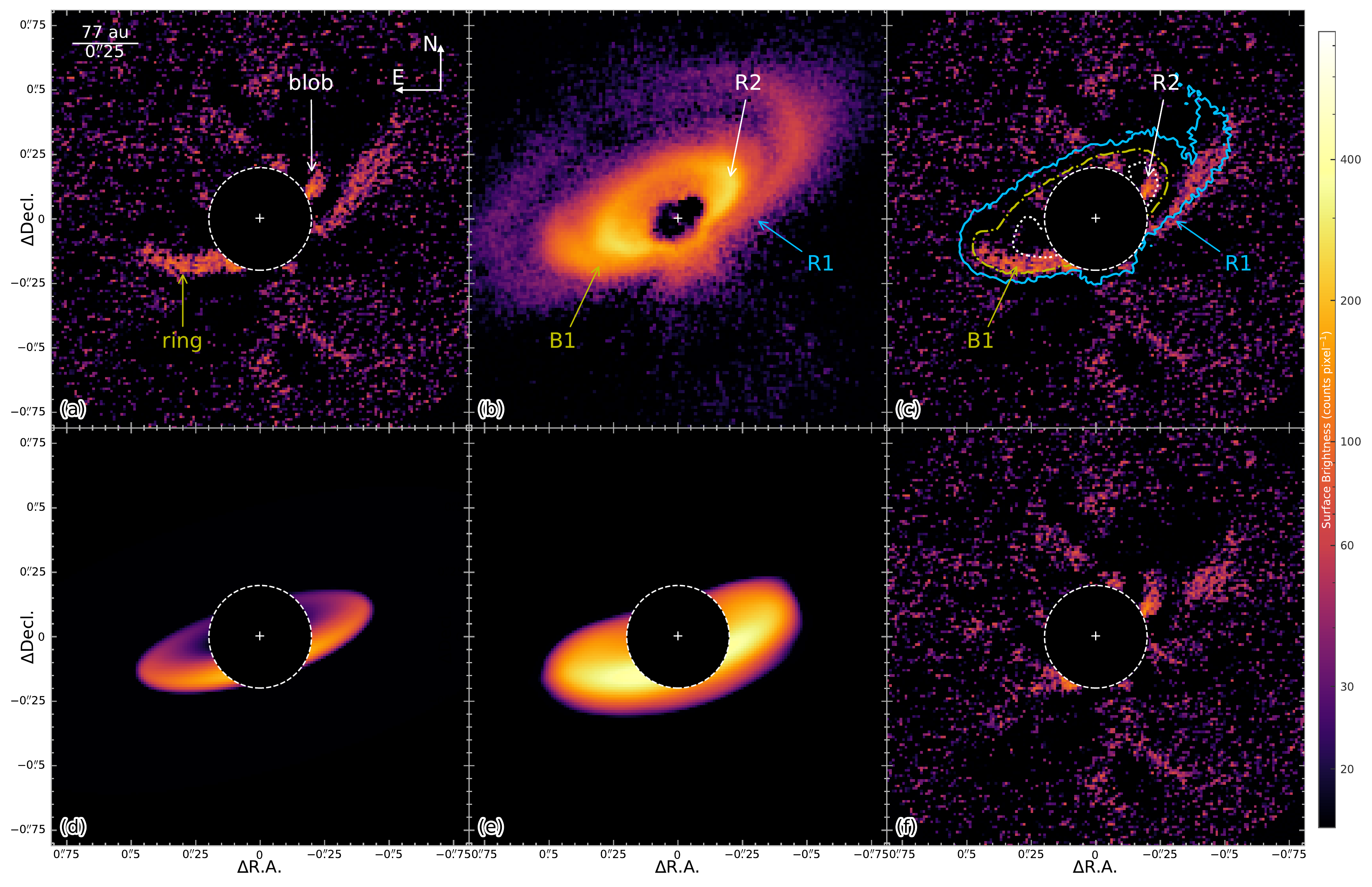}
    \caption{Keck/NIRC2 imaging and modeling of the HD~34282 system in $L'$-band. (a) Reduction result of the original exposures. (b) SPHERE \Qphi\ image, following Figure~2 of \citet{deBoer21} annotations. (c) SPHERE contours overlaid on the NIRC2 image: we confirm the existence of the B1 spiral. (d) Best-fit disk model for the NIRC2 ring. (e) Best-fit disk model with instrumentation effects taken into account (i.e., convolution, transmission) for a representative one of the 69 exposures: the actual convolution is performed for all of the exposures. (f) Reduction residuals after removing the best-fit instrumental disk model from the original exposures. Note: the images are shown in log scale, the dashed central region with a $20$~pixel radius is masked out in our data analysis; and the SPHERE data have different display limits from the NIRC2 data.
    }
    \label{fig1} 
    
    (The data used to create this figure are available in the ``anc'' folder on arXiv.)
\end{figure*}

We observed HD~34282 using Keck/NIRC2 in $L'$-band (central wavelength: $3.8~\mu$m) under program C328 (PI: G.~Ruane) on UT 2017 February 07 from 04:43 to 07:15. We use the narrow camera which has a pixel size of $9.942$~mas and a $1024{\times}1024$ pixel field of view (i.e., $10\farcs18{\times}10\farcs18$). With 73 science frames covering a parallactic angle change of $61\fdg3$, where each frame consists of 45 coadds with 1~s exposures, we have a total integration time of $3285$ s on HD~34282. During the observation, the coherence time was $\tau_0\approx0.87$~ms at $0.5~\mu$m in \citet{xuan18}, the WRF seeing\footnote{\url{http://mkwc.ifa.hawaii.edu/current/seeing/index.cgi}} was between $1\farcs32$ and $1\farcs45$, the airmass was $1.19\pm0.05$. We preprocess the data using the {\tt VIP} \citep{GomezGonzalez2017} package that is customized for NIRC2 vortex observations by performing flat-fielding, background removal, and image centering. By comparing the central pixels between the output point spread function (PSF) and an ideal model, the Strehl ratio was between $0.63$ and $0.66$ for this observation. See \citet{xuan18} for a description of both our pipeline that utilizes the {\tt VIP} package and a thorough characterization of the Keck/NIRC2 vortex coronagraph.

To reveal the extent of the disk while maximizing computational efficiency, we crop each preprocessed image to a $201{\times}201$ pixel field while discarding the last 4 exposures due to poor image quality. In the following analysis, we focus on an annular region that is between 20 pixels and 100 pixels from the star. To extract the HD~34282 system from the raw observations, we first capture the speckle features using the principal-component-analysis-based Karhunen--Lo\`eve image projection (KLIP; \citealp{soummer12}) method implemented in the {\tt DebrisDiskFM} package \citep{ren19}. To balance speckle removal and ADI self-subtraction, with the former requiring more and the latter requiring fewer KLIP components in data reduction, we use the first 6 KLIP components to remove speckles while preserving the morphology of the disk. We rotate each of the reduced 69 images to north-up and east-left, then calculate their mean to obtain the final image for the HD~34282 system, shown in Figure~\ref{fig1}a.

We also acquire the \citet{deBoer21} $J$-band observations of HD~34282 in polarized light from UT 2015 December 19 using SPHERE/IRDIS \citep[e.g.,][]{deBoer20} under European Southern Observatory (ESO) program 096.C-0248(A) (PI: J.-L.Beuzit) from the ESO Science Archive Facility. To reveal the extent of the disk, we use {\tt IRDAP} which performs polarimetric differential imaging (PDI) from \citet{vanholstein20} to reduce the IRDIS observations. In Figures~\ref{fig1}b and ~\ref{fig1}c, we show the SPHERE \Qphi\ map, and overlay the SPHERE contours (the contour levels are selected to match features in both datasets) on our NIRC2 data.

\section{Analysis}\label{sec-ana}

\subsection{Disk features}
The ADI image of HD~34282 in NIRC2 $L'$-band is comprised of a ring and a blob in the north-west direction (shown in Figure~\ref{fig1}a). The ring extends to ${\approx}0\farcs5$, and its south-east half is brighter than the north-west half by ${\approx}40\%$. The north-west blob is located outside the inner working angle of $0\farcs2$ at a position angle of $-60^\circ$ east of north.

The north-west NIRC2 blob is not a planetary signal. Comparing with the SPHERE PDI data in Figure~\ref{fig1}c, it is the inner ring component ``R2'' identified in \citet{deBoer21}. Nevertheless, we do not recover the south-east side of R2 in our NIRC2 observations; this could be explained by self-subtraction effects with ADI using KLIP. In order to recover the whole extent of R2, either reference star differential imaging \citep[e.g.,][]{ruane19}, or advanced ADI speckle removal methods \citep[e.g.,][]{ren20, pairet21, flasseur21}, are needed.

The NIRC2 ring, which has an apparent brightness asymmetry, is in fact composed of multiple features identified in the SPHERE data in \citet{deBoer21}. Specifically, the south-east half of the NIRC2 ring is a superposition of the SPHERE arm and outer ring, or the ``B1'' and ``R1'' features in \citet{deBoer21}, respectively. As evident in the SPHERE data, the south-east half of R1 is  fainter than B1, causing B1 to dominate the signal in our NIRC2 data.

\subsection{Disk forward modeling}
Observing an apparent ring-like structure in our $L'$-band data in Figure~\ref{fig1}a, we use the \citet{millarblanchaer15} code\footnote{\url{https://github.com/maxwellmb/anadisk_model}} to model the system with a ring. To take into account the observed apparent brightness asymmetry of HD~34282 evident in both the SPHERE data presented in \citet{deBoer21} and our NIRC2 data, we use the function in the updated code which can offset the ring from the assumed central star location (presented in \citealt{millarblanchaer16}). We aim to reproduce the observed ring using geometrical models to investigate the affect of disk modeling on planet detection, therefore we assume the disk is optically thin for simplicity. Thus, we do not focus on retrieving physical parameters for the observed HD~34282 ring, nor on the physical meaning of the ring center offset from the star (e.g., Table~1 of \citealp{deBoer21}). We acknowledge that such a disk model is not physically motivated, especially after the polarized light observations in \citet{deBoer21}, therefore, we only focus on the visible structure seen in the NIRC2 $L'$ data and not on its physical origins.

We use the \citet{ren21} modification of the \citet{millarblanchaer15} code, which describes the ring with a double power law \citep{augereau99} along the mid-plane and a Gaussian dispersion along the vertical axis:
\begin{equation}\label{eq1}
\rho(r, z)\propto \left[\left(\frac{r}{r_{\rm c}}\right)^{-2\alpha_{\rm in}}+\left(\frac{r}{r_{\rm c}}\right)^{-2\alpha_{\rm out}} \right]^{-\frac{1}{2}} \exp\left[-\left(\frac{z}{hr}\right)^2 \right],
\end{equation}
where $r_{\rm c}$ is the critical radius, $\alpha_{\rm in}$ and $\alpha_{\rm out}$ are the asymptotic power law indices interior and exterior to $r_{\rm c}$, and $h$ is the scale height. We use the {\tt DebrisDiskFM} framework by \citet{ren19} to explore disk parameters using {\tt emcee} \citep{emcee}. Specifically, we perform forward modeling to minimize the residuals in our reduction. We first generate a ring model using the \citet{millarblanchaer15} code assuming given ring parameters (i.e., inclination, position angle, brightness, ring center offsets along the apparent major and minor axes, critical radius, power-law indices) and forward scattering coefficient of the dust particles \citep[i.e.,][]{hg41}. We then simulate the NIRC2 instrument response on the disk model by rotating the model according to the parallactic angles for each of the $69$ exposures, convolving with the PSF, and multiplying the result by the transmission map of the NIRC2 vortex coronagraph. 

To obtain the best-fit disk model for the observed disk structure, we note that disk signals are overfit by speckle features using KLIP \citep{pueyo16}, we thus perform forward modeling using negative injection of the disk signals to minimize the residuals. Specifically, we remove the instrument's response to the disk models from the original exposures, and perform KLIP ADI reduction using $6$ components (i.e., identical KLIP parameters in original data reduction in Section~\ref{sec-obs}). The best-fit model is the one that minimizes the chi-squared residuals. We list in \ref{app-a} the best-fit parameters, with their uncertainties derived assuming the pixels are mutually independent.
By focusing on the ring regions, we obtain the best-fit disk model in Figure~\ref{fig1}d, and the corresponding residual map is presented in Figure~\ref{fig1}f.

Due to the existence of the ring brightness asymmetry in NIRC2 and the two-ringed architecture in SPHERE, we have performed additional model fitting not shown in Figure~\ref{fig1}. We have attempted to fit the two halves (i.e., north-west and south-east) of the NIRC2 ring with two half-rings both centering at the star. The two half-rings drift towards nearly edge-on, inconsistent with observations of this system, which suggests they are likely offset from the star.  We have also tried to fit the north-west blob in NIRC2 with a smaller ring, as motivated by the existence of R2 in the SPHERE observations. We cannot find a model that fits for this structure, since such a ring requires a counterpart in the south-east (e.g., Figure~\ref{fig1}c) that is missing in the NIRC2 observations. We thus adopt an offset ring to describe the NIRC2 data to minimize the number of free parameters in our model.

\begin{figure}[htb!]
\centerfloat
\includegraphics[width=.49\textwidth]{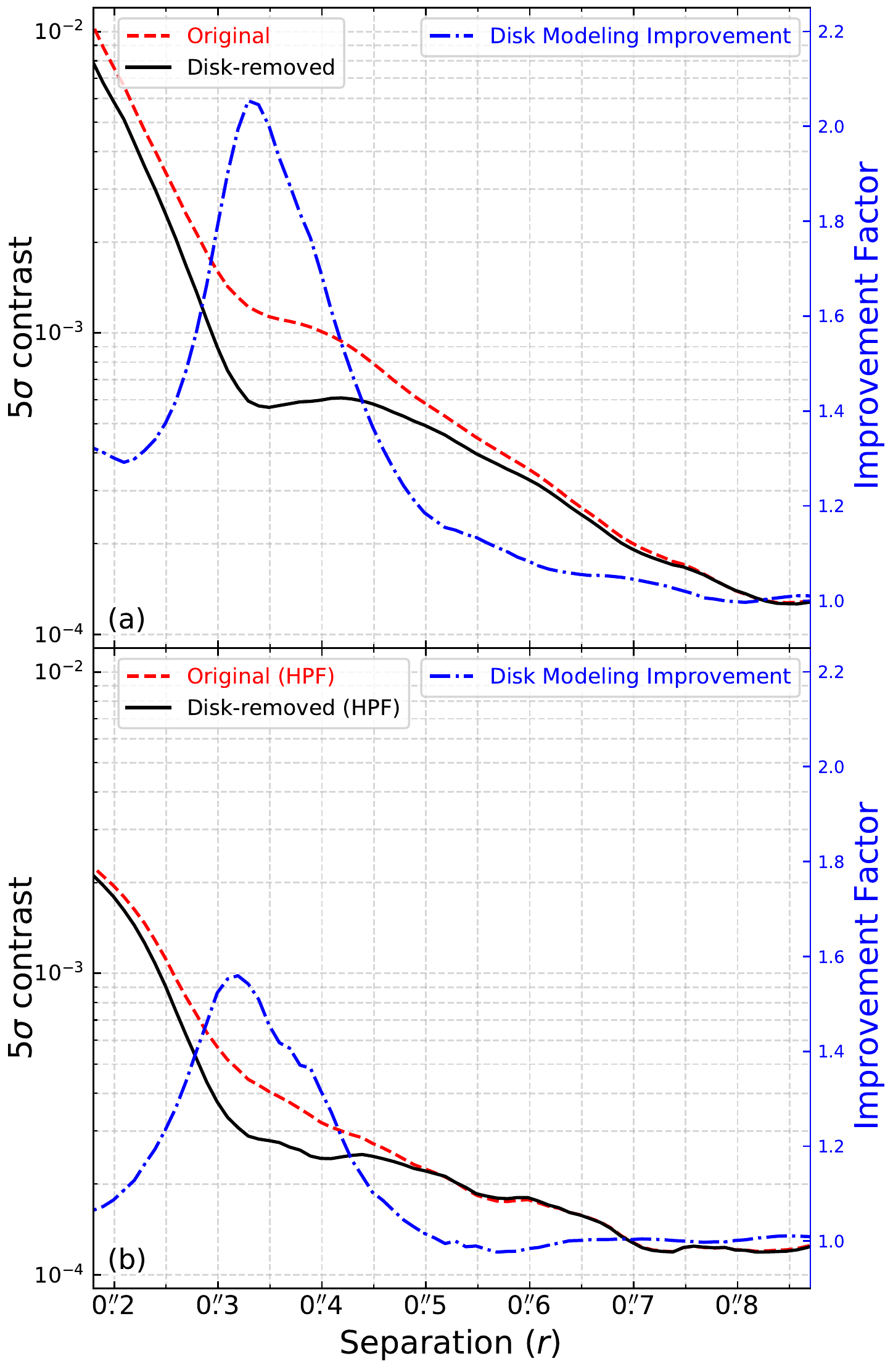}
\caption{5$\sigma$ detection limit of point sources as a function of radial separation for our observations. The red dashed line is the contrast curve for the original data, and black solid for the data with the best-fit disk model removed. By dividing the two contrast curves, see blue dash-dotted line, we obtain a better overall detection limit. For the disk region, when removing the disk, we are more sensitive by a factor of $\sim$2 over the original data in panel (a), and $\sim$1.6 for the high-pass filtered data in panel (b).}
\label{contrast} 
\end{figure}

\subsection{$L'$-band magnitude of HD~34282}

The \textit{WISE} $W1$-band covers a fraction of the Keck/NIRC2 $L'$-band wavelengths, we thus use the former to obtain the $L'$ magnitude for HD~34282. We first generate a high-resolution stellar spectrum using a PHOENIX stellar model from \citet{Husser13}, and interpolate within the grid to match the stellar parameters from \citet{Mer04}. In order to match the model spectrum with the observed \textit{WISE} $W1$ value of $7.072\pm0.016$ (CatWISE2020; \citealp{Marocco21}), we scale the resulting model flux when integrated across the $W1$ bandpass \citep{Rodrigo12,Rodrigo20}. We then take the resulting scaled spectrum, account for the transmission through the atmosphere, and integrate across the Keck/NIRC2 $L'$ filter profile \citep{Rodrigo12,Rodrigo20} to obtain an $L'$ magnitude of $7.729_{-0.017}^{+0.015}$.

\subsection{Detection limits}
We obtain the detection sensitivity to planets by injecting point sources into the observations and calculating their signal-to-noise (S/N) values after KLIP ADI reduction. To investigate the affect of disk modeling on point source detection, we calculate the contrast before and after removing the best-fit model in Figure~\ref{fig1}d from the observations. We calculate 1D contrast curves with radial distance from the host star using {\tt VIP} which utilizes fake companion injection and recovery when determining the contrast achieved at different radial distances. We compare the contrast achieved in the image where we remove the best-fitting disk to that in which we do not in Figure~\ref{contrast}. Comparing the two contrast curves, we find a marked improvement in the contrast achieved at the radial separations where the disk is visible. We also test the affect of utilizing a high-pass filter (HPF) to remove the disk structure, which in essence is treating the disk like noise. We find that an improvement is seen for both the original data and the data in which we use a HPF\footnote{We smooth each raw exposure with a Gaussian that is 2 times the full-width at half-maximum of the PSF, and remove it from the raw exposure.}.

\begin{figure}[htb!]
\includegraphics[width=.475\textwidth]{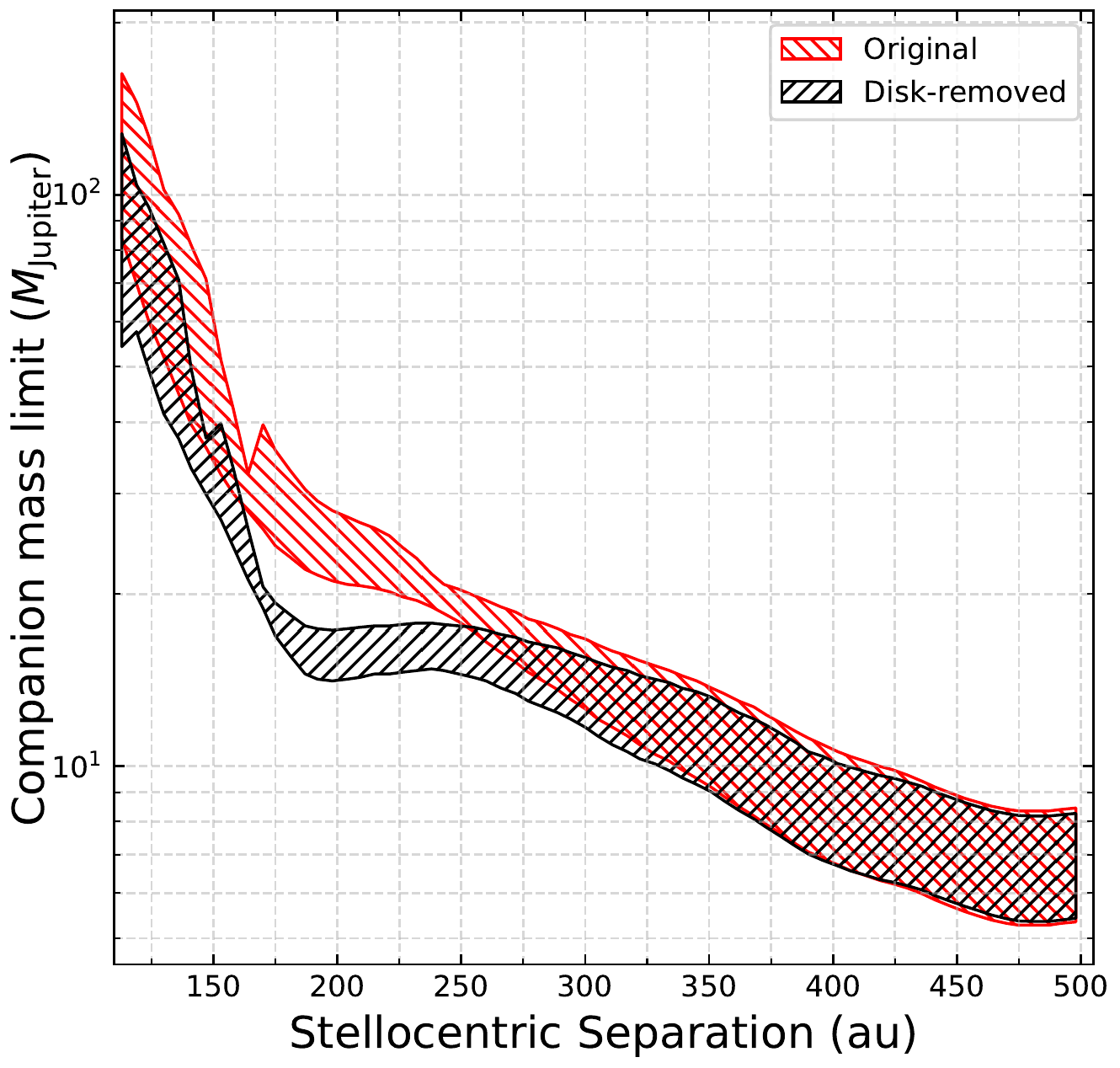}
\caption{Mass limits using the AMES-Cond models for our observations with the disk present (in red) and after removing the disk (in black) as a function of separation in au (after accounting for inclination-based projection affects). The width of the shaded regions account for uncertainties in the distance, age, and magnitude of the host star. We are sensitive to planets $\sim$10~$M_{\rm Jupiter}$ smaller after removing the best fitting disk model in the region where the disk is present. Note: the minimum stellocentric separation presented here is outside the coronagraph to avoid possible coronagraphic occultation; future analysis at shorter separations should take the occultation probability into account.}
\label{mass} 
\end{figure}

While using a HPF allows us to achieve better contrast than not using a HPF, the data with the HPF does degrade the quality of the extended disk structure seen this system, therefore, we focus on the data without the HPF in order to highlight the affect that disk modeling has. For the data without the HPF, we then utilize our 1D contrast curves and the AMES-Cond models \citep{Baraffe2003Evolutionary209458} to determine what mass planets we are sensitive to with our improved contrast limits after removing the disk. The AMES-Cond evolutionary models predict how luminous an object will be given a mass and age. We use our $L'$ stellar magnitude, our contrast limits, and the published age for this system (6.4$^{+1.9}_{-2.6}$ Myr; \citealt{vanderPlas17}) to determine expected mass limits in the disk plane based on disk parameters in \citet{vanderPlas17}. We interpolate within the AMES-Cond model grid and determine the mass corresponding to our contrast with radial separation in Figure~\ref{mass}.

We can reach a typical sensitivity of ${\sim}10~M_{\rm Jupiter}$ mass assuming the AMES-Cond models. However, our sensitivity is sub-optimal in the region where the disk signal resides at ${\sim}200$~au, as has been observed in Figure~6 of \citet{deBoer21}. Utilizing disk modeling, we are able to reach a sensitivity of ${\sim}15~M_{\rm Jupiter}$ at ${\sim}200$~au, a factor of ${\sim}2$ better in comparison with the original reduction. As a result, we are able to achieve better sensitivity by performing disk modeling, allowing us to be more sensitive to smaller planets, ultimately possibly allowing for better constraints on the occurrence rate of less massive planets.

\section{Summary}\label{sec-sum}
We model the $L'$ observation of the single-armed spiral disk HD~34282 using an offset geometric disk model. We show that even using a non-physical simplified model allows for better achieved sensitivity over not modeling the disk. Therefore, we recommend modeling the disks in scattered light systems to increase the sensitivity to less massive planets. In fact, the importance of modeling observed disks  is demonstrated in \citet{wang20}, where the luminosity of PDS~70c was only evident after disk modeling. In addition to the established two steps of planet imaging (instrumentation and observation design, and post-processing for speckle noise removal), our study here suggests that a third step is necessary in cases where we detect disks in scattered light -- performing disk modeling to disentangle non-planetary astrophysical signals to reveal possibly obscured planets.

We caution that for the purpose of targeted planet searches, one should mask out expected planet regions during the disk modeling process as in \citet{wang20} for PDS~70c. In our study of NIRC2 imaging of HD~34282, one additional caveat is that possible planets can be fit out by the offset ring model. Nevertheless, the simulations in \citet{dong17} suggest that planets inducing a single spiral arm can be $0.1$--$1~M_{\rm Jupiter}$. Such a planet is expect to be not detectable even after us removing the disk signal following the \citet{wang20} approach. Despite this, we note that a variation of the \citet{wang20} approach -- specifically, masking out different regions while disk modeling for an exhaustive search of unknown planets -- could still be necessary for other systems.

We also caution that physically motivated disk models satisfying dynamical constraints are preferred in order to characterize the scattered light signals while modeling. Our simple assumptions in the models presented herein show a first attempt at subtracting disk signals in order to improve planet detection capabilities. Future studies, including those along the lines of, e.g., \citet{wolff17} and \citet{ villenave19}, to characterize the dust properties in the HD~34282 disk, are needed to refine the models. 

\begin{acknowledgements}
We thank the anonymous referee for their comments that improved this Letter. This research is partially supported by NASA ROSES XRP, award 80NSSC19K0294. We thank Jean-Baptiste Ruffio, Jason Wang, Christian Ginski, and Myriam Benisty for discussions on calculating contrast curves. R.D.~acknowledges financial support provided by the Natural Sciences and Engineering Research Council of Canada through a Discovery Grant, as well as the Alfred P.~Sloan Foundation through a Sloan Research Fellowship. Some of the data presented herein were obtained at the W.~M.~Keck Observatory, which is operated as a scientific partnership among the California Institute of Technology, the University of California and the National Aeronautics and Space Administration. The Observatory was made possible by the generous financial support of the W.~M.~Keck Foundation. The authors wish to recognize and acknowledge the very significant cultural role and reverence that the summit of Maunakea has always had within the indigenous Hawaiian community.  We are most fortunate to have the opportunity to conduct observations from this mountain. Part of the computations presented here were conducted in the Resnick High Performance Computing Center, a facility supported by Resnick Sustainability Institute at the California Institute of Technology.  Based on observations collected at the European Organisation for Astronomical Research in the Southern Hemisphere under ESO programme 096.C-0248(A). This publication makes use of data products from the \textit{Wide-field Infrared Survey Explorer}, which is a joint project of the University of California, Los Angeles, and the Jet Propulsion Laboratory/California Institute of Technology, funded by the National Aeronautics and Space Administration. This research has made use of the SVO Filter Profile Service (\url{http://svo2.cab.inta-csic.es/theory/fps/}) supported from the Spanish MINECO through grant AYA2017-84089.
\end{acknowledgements}

\facility{Keck II (NIRC2), VLT:Melipal (SPHERE)}

\software{{\tt DebrisDiskFM} \citep{ren19}, {\tt emcee} \citep{emcee}, {\tt IRDAP} \citep{vanholstein20}, {\tt VIP} \citep{GomezGonzalez2017}}

\appendix

\section{Disk Modeling Parameters}\label{app-a}
We list the best-fit parameters and their meanings in our disk models in Table~\ref{tab1}. We have not adopted a physically motivated model to describe the observed ring, but analytically modeled the observed distribution of light assuming the disk is optically thin. Due to the unphysical nature of our disk model, we do not further discuss the physical implications of these parameters.

\bibliography{refs}

\begin{rotatepage}
\begin{rotatetable}
\begin{deluxetable*}{ccccl}
\tablecaption{Posterior values for the HD~34282 ring parameters in Keck/NIRC2 $L'$-band \label{tab1}}
\tablehead{ 
\colhead{\multirow{2}{*}{Parameter}}  & \colhead{Maximum} & \colhead{{$50$th $\pm34$th}} & \colhead{\multirow{2}{*}{Unit}} & \colhead{\multirow{2}{*}{Parameter Meaning}}\\
 & \colhead{Likelihood} & \colhead{Percentiles} &\\
 \colhead{(1)} & \colhead{(2)} & \colhead{(3)} & \colhead{(4)} & \colhead{(5)}
 }
\decimals
\startdata 
$g$ & $0.38$& $0.40_{-0.03}^{+0.05}$ & & Forward scattering parameter in the \citet{hg41} phase function.\\
$\theta_{\rm inc}$ & $72.3$ & $-73_{-2}^{+2}$ &  degree &Inclination of the ring, measured from face-on.\\
$\theta_{\rm PA}$ & $-74.1$ & $-73.8_{-1.5}^{+1.4}$ &  degree& Position angle of the apparent major axis of the ring, measured east of north.\\
$r_{\rm c}$ & $140$ & $135_{-9}^{+9}$ &  au& Critical radius of the dust distribution for the ring model, see Equation~\eqref{eq1}.\\
$\alpha_{\rm in}$ & $2.1$ & $6_{-3}^{+8}$ &  & Asymptotic power law index of dust distribution for $r \ll r_{\rm c}$, see Equation~\eqref{eq1}.\\
$\alpha_{\rm out}$ & $-23.1$ & $-19_{-8}^{+8}$ &  & Asymptotic power law index of dust distribution for $r \gg r_{\rm c}$, see Equation~\eqref{eq1}.\\
$\Delta X_1$ & $2.7$ & $2_{-14}^{+11}$ &  au & Ring center offset along the disk major axis, positive is towards southeast \citep{millarblanchaer16}.\\
$\Delta X_2$ & $8.5$ & $7_{-5}^{+5}$ &  au & Ring center offset along the disk minor axis, positive is towards disk backside \citep{millarblanchaer16}.\\
\enddata
\tablecomments{Column 1: disk parameters of interest. Column 2: best-fit parameters used to generate the best-fit disk model. Column 3: $50$th$\pm34$th percentiles assuming independent pixels. Column 4: parameter units. Column 5: meaning of parameters.
}
\end{deluxetable*}
\end{rotatetable}
\end{rotatepage}

\end{CJK*}
\end{document}